\documentstyle[12pt]{article}
\newcounter{mycount}

\newcommand{\G}{{\fr G}}
\newcommand{\be}{\begin{equation}}
\newcommand{\R}{{\bf R}}
\newcommand{\p}{\partial}
\newcommand{\ee}{\end{equation}}
\newcommand{\bee}{\begin{eqnarray}}
\newcommand{\eee}{\end{eqnarray}}
\newcommand\nn{\nonumber \\}
\newcommand\als{associative algebras }
\newcommand{\ie}{$i.e.\ $}

\font\frtnfr=eufm10   scaled\magstep2
\font\twlfr=eufm10
\font\tenfr=eufm10

\newfam\frfam
\textfont\frfam=\frtnfr
\scriptfont\frfam=\twlfr
\scriptscriptfont\frfam=\tenfr
\def\fr{\fam\frfam}

\font\frtnopen=msbm10  scaled\magstep2
\font\twlopen=msbm10
\font\tenopen=msbm10

\newfam\openfam
\textfont\openfam=\frtnopen
\scriptfont\openfam=\twlopen
\scriptscriptfont\openfam=\tenopen

\font\frtnsf = cmss12 scaled\magstep1
\font\twlsf = cmss10
\font\tensf = cmss9

\newfam\Scfam
\textfont\Scfam = \frtnsf
\scriptfont\Scfam = \twlsf
\scriptscriptfont\Scfam = \tensf

\begin{document}
\renewcommand{\theequation}{\arabic{equation}}
\bibliographystyle{nphys}
\setcounter{equation}{0}

\centerline{\Large\bf Supertraces on the algebra of observables of the}
\centerline{\Large\bf rational Calogero model based on the classical}
\centerline{\Large\bf root system}
\vskip 7mm

\centerline{S.E.~Konstein}

\vskip 3mm
\noindent
\centerline{I.E.Tamm Department of Theoretical Physics,
P. N. Lebedev Physical Institute,} \\
\centerline {117924
Leninsky Prospect 53, Moscow, Russia.}
\vskip 3mm \noindent
\vskip 7mm
\renewcommand{\theequation}{\arabic{equation}}
\setcounter{equation}{0}
\begin{abstract}
We find a complete  set of supertraces on the algebras $H_{W(\R)}(\nu)$,
the algebra of observables of the rational Calogero model with harmonic
interaction based on the classical root systems $\R$ of $B_N$, $C_N$ and
$D_N$ types.  These results extend the results known for the case $A_{N-1}$.
It is shown that $H_{W(\R)}(\nu)$  admits $q(\R)$ independent supertraces
where $q(B_N)=q(C_N)$ is a number of partitions of $N$ into a sum of positive
integers and $q(D_N)$ is a number of partitions of $N$ into a sum of positive
integers with even number of  even integers.
\end{abstract}

\section{Introduction}

In this paper we continue to investigate some properties of the associative
algebras that were shown in \cite{1,2} to underly the rational Calogero model
\cite{4}.  We extend the results obtained in \cite{kv} to the \als that
underly the rational Calogero model associated with the root systems
\cite{op} of the classical Lie algebras $B_N$, $C_N$ and $D_N$.

Let $H_G(\nu)$ be the associative algebra generated by $N$ pairs of
(deformed) Heisenberg-Weyl oscillators $a_i^\alpha$ which constitute the
basises in the pair of $N$-dimensional subspaces $a_i^\alpha \in {\cal
H}_N^\alpha$ and group algebra $\G$ of some finite group $G$ with the
following properties:

i) $H_G(\nu)$ possesses the parity $\pi$: $\pi (a_i^\alpha)=1$, $\pi(g)=0$
$\forall g\in{G}$

ii) $g a^\alpha_i = \sum_j T_{ij}(g^{-1}) a^\alpha_j g$, $\forall g \in G$
and for all $\alpha=0,1$, $i=1,...,N$, $ a^\alpha_i \in {\cal H}_N^\alpha$,
where the orthogonal matrices $T_{ij}(g) $ realize $N$-dimensional
representation of group $G$.

iii) $[a^\alpha_i, a^\beta_j] = \epsilon^{\alpha\beta} A_{ij}$, where
$\epsilon^{\alpha\beta}$ is antisymmetric tensor, $\epsilon^{01}=1$, and
$A_{ij}=A_{ji}\in \G$ depend on one or more parameters $\nu$.

To formulate the next property let us introduce the subspaces ${\cal
E}^\alpha (g) \subset {\cal H}_N$ as \be {\cal E}^\alpha (g) =\{h\in {\cal
H}_N^\alpha:\quad gh=-hg \}\,, \ee and the grading $E$ on ${\fr G}$, \be
E(g)=dim\,{\cal E}^\alpha(g).  \ee

iiii) $E({\cal P}([h_0,\,h_1])g)=E(g)-1$ $\forall g\in G$, $\forall
h_\alpha\in {\cal E}^\alpha(g)$. Here the notation ${\cal P}(g)$ introduced
for the projector $\G \rightarrow \G$ defined as ${\cal P}(\sum_i \alpha_i
g_i)=\sum_{i:\, g_i \neq {\bf 1}} \alpha_i g_i$ for $g_i\in G$ and any
constants $\alpha_i$.

v) Every element in $H_G(\nu)$ is a polynomial of $a_i^\alpha$.

We define the supertrace as a linear complex-valued function $str(\cdot)$ on
the algebra $H_G(\nu)$ such that \bee\label{scom}
str(fg)&=&(-1)^{\pi(f)\pi(g)}str(gf)\quad \forall f,g \in \mbox{$H_G(\nu)$}.
\eee

On the example of the case $G=S_N$ the following theorem is proved in
\cite{kv}.

{\it {\bf Theorem 1.} If the superalgebra $H_G(\nu)$ possesses the properties
i) - v) then the number of independent supertraces on $H_G(\nu)$ is equal
to the number of the supertraces on the algebra $\G$, satisfying the
equations \bee\label{GLC1} str([h_0,\,h_1]g)=0\qquad \forall g\in G \mbox{
with }E(g)\neq 0 \mbox{ and }\forall h_\alpha \in {\cal E}^\alpha(g).  \eee }

In this paper we apply this theorem to the case where $G$ is Weyl group $W(\R
)$ of the root system $\R $ of classical Lie algebra, $A_{N-1}$, $B_N$,
$C_N$, and $D_N$ types.  We will consider three (because $W(B_N)=W(C_N)$)
different cases simultaneously. Note that the algebra $SH_N(\nu)$ considered
in \cite{5, kv} is the algebra $H_{W(A_{N-1})}(\nu) = H_{S_N}(\nu)$ in our
notations.

Consider 3-parametric deformation $H_G(\nu)$ of the associative
Heisenberg-Weyl algebra of polynomials 
of $N$ pairs of oscillators.  This algebra is
generated by $N$ pairs of generating elements $a^\alpha_i$,  $\alpha=0,1$,
and by the reflections $K_{ij}$ and $R_i$, $i,j=1,\,...,\,N$, satisfying the
following relations \footnote{In this paper repeated Latin indices
$i,j,k,\ldots$ do not imply summation.}:  \bee\label{KK} K_{ij}=K_{ji},\
K_{ij}\,K_{ij}=1,\  K_{ij}K_{jl}= K_{jl}K_{li}= K_{li}K_{ij}\qquad \mbox{ for
$i\neq j \neq l\neq i$,} \nn K_{ij}\,K_{kl}=K_{kl}\,K_{ij}\qquad \mbox{ if
$i,\,j,\,k,\,l$ are pairwise different,} \eee \bee\label{RR} R_i R_j = R_j
R_i\quad \forall i,j\,, \qquad R_i R_i =1,\quad \forall i\,, \eee
\bee\label{KR}\label{RK} K_{ij}R_j=R_iK_{ij}, \quad K_{ij}R_k=R_k
K_{ij}\qquad \mbox{for } i\neq j\neq k \neq i, \eee \bee\label{RA}\label{AR}
R_i a^\alpha_i=-a^\alpha_iR_i ,\qquad R_i a^\alpha_j= a^\alpha_jR_i \mbox{
for }i\neq j, \eee \bee\label{KA}\label{AK} K_{ij} a^\alpha_j=
a^\alpha_iK_{ij},\qquad K_{ij} a^\alpha_k= a^\alpha_kK_{ij} \mbox{ for }i\neq
j\neq k \neq i, \eee \bee\label{AA} \left [ a^\alpha_i\,,\,a^\beta_j \right
]= \epsilon^{\alpha\,\beta} A_{ij}\,, \eee where $\ \epsilon^{\alpha\,\beta}$
=$- \epsilon^{\beta \,\alpha},$ $\ \epsilon^{0\,1}=1$, and \bee\label{A}
A_{ij}&=&\delta_{ij} +\nu_0\tilde{A}^0_{ij} +\nu_1\tilde{A}^1_{ij}
+\nu_2\tilde{A}^2_{ij}\,,\nn \tilde{A}^0_{ij}&=&\delta_{ij}\sum_{l=1,\,l\neq
i}^N K_{il}- \delta_{i\neq j} K_{ij}\,,\nn
\tilde{A}^1_{ij}&=&\delta_{ij}R_i\,,\nn
\tilde{A}^2_{ij}&=&\delta_{ij}\sum_{l\neq i} K_{il}R_iR_l+ \delta_{i\neq
j}K_{ij}R_iR_j\,.  \eee

The commutation relations defined in such a way are selfconsistent when one
of the following conditions takes place:

\noindent {\bf A)} $\nu_1=\nu_2=0$, and generating elements $R_i$ are
excluded,\\ {\bf B,C)} $\nu_0=\nu_2$, \\ {\bf D)} $\nu_0=\nu_2$, $\nu_1=0$,
and every monomial in $H_G(\nu)$ contains even number of reflections $R_i$.

The reflections $K_{ij}$ and $R_i$ determine Weyl group of the root systems
$A_{N-1}$, $B_N$, $C_N$ and $D_N$ correspondingly to these cases and the
operators $a^\alpha_i$ have the presentation \bee\label{pres} a^\alpha_i= x_i
+(-1)^\alpha D_i(x) \eee where $D_i(x)$ are Dunkl's differential-difference
operators \cite{6} connected with the corresponding root systems,
\bee\label{dunkl} D_i= \frac \p {\p x_i}+ \nu_1\frac 1 {x_i}(1-R_i)+
\sum_{l\neq i}^N \left(\nu_0 \frac {1} {x_i-x_l}(1-K_{il}) +\nu_2 \frac {1}
{x_i+x_l}(1-K_{il}R_iR_l)\right), \eee and satisfying the condition \bee
\left[D_i,\,D_j\right]=0 \qquad  \forall i,j \eee for values of $\nu$-s
listed above.  \footnote {In (\ref{dunkl}) the reflections $K_{ij}$ and $R_i$
act on the space of coordinates $x_i$}

The Hamiltonian of Calogero model associated with corresponding root system
\cite{op} is identified with second-order differential operator $H=\frac 1 2
\sum_{i=1}^N \left\{a^0_i,\, a^1_i\right\}$.  The operators $a_i^\alpha$
serve as generalized oscillators underlying the Calogero problem and allow
one \cite{2} to construct wave functions via the standard Fock procedure with
the Fock vacuum $ |0\rangle$ such that $a_i^0 |0\rangle$=0.

To know the supertraces is useful in various respects.  One of the most
important is that they define multilinear invariant forms \be\label{bil}
str(a_1 a_2\,...\,a_n), \ee what allows for example to construct the
lagrangians for dynamical theories based on these algebras.  Another useful
property is that since null vectors of any invariant bilinear form span a
both-side ideal of the algebra, this gives a powerful device for
investigating ideals which decouple from everything under the supertrace
operation as it happens in $SH_2 (\nu )$ for half-integer $\nu$  \cite{14}.

An important motivation for the analysis of the supertraces of $H_{W({\bf
R})}(\nu)$ is due to its deep relationship with the analysis of the
representations of this algebra, which in its turn gets applications to the
analysis of the wave functions of the Calogero models. For example, given
representation of $H_{W({\bf R})}(\nu)$ , one can speculate that it induces
some supertrace on this algebra as (appropriately regularized) supertrace of
(infinite) representation matrices. When the corresponding bilinear form
(\ref{bil}) degenerates this would imply that the representation becomes
reducible.

For almost all superalgebras considered in this paper the situation is very
interesting since almost all of them admit more than one independent
supertrace.  For finite dimensional algebras the existence of several
supertraces means that they have both-side ideals, but for infinite
dimensional algebras under consideration the existence of ideals is still an
open problem.

Below, in the section \ref{-1} we prove the lemma, that ensures the existence
of the supertraces on $H_{W({\bf R})}(\nu)$ and in the section \ref{Q_N} we
calculate the number of supertraces using the theorem proved in Appendix.

\section{1-dimensional representations of the elements of the Weyl groups of
the classical root systems}\label{-1}

In this section we show that the property iiii) is satisfied for $H_G(\nu)$
if the group $G$ is the Weyl group of the classical root system, and is
generated by the reflections $K_{ij}$ and $R_i$.

Every element $g\in G$ for $G= W(A_{N-1})$, $G= W(B_N)$ $G= W(C_N)$ and $G=
W(D_N)$ can be presented in the form $g=\sigma \prod_{i\in M_g} R_{i}$ where
the permutation $\sigma \in S_N$, the symmetric group, and $M_g$ is some
subset of indices $1,...,N$.  It is well known that every permutation
$\sigma\in S_N$ defines some partition of the set of indices $1,\,...\,,\,N$
to the subsets $C_1,\,...\,,C_{t_\sigma}$ in such a way that it can be
presented as a product $\sigma=\prod_{{\fr m}=1}^{t_\sigma}c_{\fr m}$ of the
commuting cycles $c_{\fr m}$ acting on the subsets $C_{\fr m}$
correspondingly.

Introduce the elements \bee r_{\fr m}=\prod_{i\in M_g\cap C_{\fr m}} R_i.
\eee Then $g$ can be presented as \bee\label{dec} g=\prod_{{\fr
m}=1}^{t_\sigma}\hat c_{\fr m} \mbox{ where $\hat c_{\fr m}=c_{\fr m}r_{\fr
m}$}.  \eee We call these elements $\hat c_{\fr m}$ as cycles.

Given element $g\in G$, we introduce a new set of basis elements ${\fr
B}_g$=$\{b^I\}$ instead of $\{a^\alpha_i\}$ in the following way.  For every
cycle $\hat c_{\fr m}$ in the decomposition (\ref{dec}) let us fix some index
$l_{\fr m}$, that belongs to the subset $C_{\fr m}$ associated with the cycle
$\hat c_{\fr m}$.  The basis elements $b^\alpha_{{\fr m}j}$,
$j=1,\,...\,,|C_{\fr m}|$, that realize 1-dimensional representations of the
commutative cyclic group generated by $\hat c_{\fr m}$, have the form \be
\label {a} b^\alpha_{{\fr m}j} =\frac 1 {\sqrt{|C_{\fr m}|}} \sum
_{k=1}^{|C_{\fr m} |} (\lambda_{\fr m})^{jk} \hat c_{\fr
m}^{-k}a^{\alpha}_{l_{\fr m}} \hat c_{\fr m}^{k}\,, \ee where
\bee\label{lambda} \lambda_{\fr m}=exp\left(\frac{1+(|\hat c_{\fr m}| +
|C_{\fr m}|)_{mod\, 2}} {|C_{\fr m}|}\pi i\right).  \eee Here the notation
$|C_{\fr m}|$ denotes the number of elements in the subset $C_{\fr m}$, and
$|\hat c_{\fr m}|$ is the length of the cycle $\hat c_{\fr m}$.  The length
$|g|$ of some element $g\in G$ is the minimal number of the reflections whose
product is $g$ \footnote{In \cite {kv} we used the definition of the length
equal to $|\hat c_{\fr m}|+1$.}.

{}From 
the definition (\ref{a}) it follows that \be\label{eig} \hat c_{\fr m}
b^\alpha_{{\fr m}j} = (\lambda_{\fr m})^j b^\alpha_{{\fr m}j} \hat c_{\fr
m}\,, \ee \be\label{eignext} \hat c_{\fr m} b^\alpha_{{\fr n}j} =
 b^\alpha_{{\fr n}j} \hat c_{\fr m}\mbox{ , for } {\fr n}\neq {\fr m}\,, \ee
and therefore \be\label{eigs} g b^\alpha_{{\fr m}j} = (\lambda_{\fr m}
)^jb^\alpha_{{\fr m}j} g\,.  \ee

Now one can easily deduce that for every cycle $\hat c_{\fr m}$ with even
length the matrix $T_{ij}(\hat c_{\fr m})$  has no eigenvalues equal to $-1$
while for the cycle with odd length it has precisely 1 eigenvalue $-1$. As a
consequence the function $E(g)$ is the number of odd cycles in the
decomposition (\ref{dec}) of $g$.

In what follows, instead of writing  $b^\alpha_{{\fr m}j}$ we use the
notation $b^I$ with the label $I$ accounting for the full information about
the index $\alpha$, the index ${\fr m}$ enumerating cycles in (\ref{dec}),
and the index $j$ that enumerates various elements $b^\alpha_{{\fr m}j}$
related to the cycle $\hat c_{\fr m}$, \ie $I$ ($I = 1,\,...,\,2N$)
enumerates all possible triples $\{\alpha,{\fr m},j\}$.  We denote the index
$\alpha$, the cycle, the subset of indices and the eigenvalue in (\ref{eig})
corresponding to some fixed index $I$ as $\alpha (I)$, $ c(I)$, $C(I)$ and
$\lambda_I=(\lambda_{\fr m})^j$, respectively.  The notation $g(I)=g_0$
implies that $b^I \in {\fr B}_{g_0}$. ${\fr B}_{\bf 1}$ is the original basis
of the generating elements $a_i^\alpha$ (here ${\bf 1}\in G$ is the unit
element).

Let ${\fr M}(g)$ be the matrix that maps ${\fr B}_{\bf 1}\longrightarrow {\fr
B}_g$ in accordance with (\ref{a}), \bee\label{frm} b^I=\sum_{i,\alpha} {\fr
M}_{i\alpha}^I(g)\, a^\alpha_i\,.  \eee Obviously this mapping is invertible.
Using the matrix notations one can rewrite (\ref{eigs}) as \bee\label{eigmat}
g b^I g^{-1}=\sum_{J=1}^{2N} \Lambda^I_J(g)\, b^J\,,\ \ \forall b^I \in {\fr
 B}_g\,, \eee where $\Lambda_I^J(g)=\delta_I^J \lambda_I$.

The commutation relations for the generating elements $b^I$ follow from
(\ref{AA}) and (\ref{A}) \be\label{f} \left [ b^I,\,b^J\right]= F^{IJ}= {\cal
C}^{IJ}+ \epsilon^{\alpha (I)\alpha (J)}B^{IJ}\,, \ee where $g(I)=g(J)$,
\be\label{calC} {\cal C}^{IJ}=\epsilon^{\alpha (I) \alpha (J)} \delta_{c(I)
c(J)} \delta_{\lambda_I \lambda_J^{-1}} \ee and \be \label{struc}
B^{IJ}=\sum_{i,j}{\fr M}^I_{i\,0} (\sigma ) {\fr M}^J_{j\,1} (\sigma )
A_{ij},\qquad \epsilon^{\alpha (I)\alpha (J)}B^{IJ}={\cal P} \left(\left [
b^I,\,b^J\right]\right).  \ee

The indices $I,J$ are raised and lowered with the aid of the symplectic form
$ {\cal C}^{IJ}$ \be\label{rise} b^I=\sum_J{\cal C}^{IJ}b_J\,,\qquad
b_I=\sum_Jb^J {\cal C}_{JI}\,; \qquad \sum_M{\cal C}_{IM}{\cal
C}^{MJ}=-\delta_I^J\,.  \ee Note that the elements $b^I$ are normalized in
(\ref{a}) in such a way that the $\nu$-independent part in (\ref{f}) has the
form (\ref{calC}).

Now we can prove the existence of the supertraces for the cases $B_N$, $C_N$
and $D_N$ using {\it Theorem 1}.  For this purpose it is sufficient to prove
the next lemma (property iiii) ).

\noindent {\it {\bf Lemma 1.}If $\lambda_I=\lambda_J=-1$, where $g(I)=g(J)=g$
then  $E(B^{IJ}g) = E(g)-1$ .}

The proof is based on the following simple facts from the theory of the
symmetric group:

\noindent {\it{\bf Proposition 1.} Let $c_1$ and $c_2$ be two distinct cycles
in the decomposition of some permutation from $S_N$.  Let indices $i_1$ and
$i_2$ belong to the subsets of indices associated with the cycles $c_1$ and
$c_2$, respectively. Then the permutation $c= K_{i_1\, i_2}c_1 c_2 $ is a
cycle of length $|c|= |c_1| + |c_2|+1 $.}

\noindent {\it{\bf Proposition 2.} Given cyclic permutation $c \in S_N$, let
$i\neq j$ be two indices such that $c^k (i) = j$, where $k$ is some positive
integer, $k<|c|$.  Then $K_{ij}c  = c_1 c_2 $ where $c_{1,2}$ are some
non-coinciding mutually commuting cycles such that $|c_1|=k-1$ and $|c_2|=
|c|-k$.}

To prove {\it Lemma 1} let us first consider the case $c(I)=c(J)$.  For the
definiteness consider the odd cycle \bee\label{odd} c(I)=c(J)= K_{12} K_{23}
\,...\,K_{(p-1)p} R_{i_1} R_{i_2}\,...\, R_{i_l} \eee with $1\leq
i_1<i_2<\,...\,<i_l\leq p$ and odd $p-1+l$ \\ and \bee\label{cycle} b^P=
\frac 1 {\sqrt p}\sum_{j=1}^p (-1)^j (-1)^{\Delta(j)} a^{\alpha(P)}_j,\qquad
P=I,J, \eee where \bee\label{Delta} \Delta(j)=\sum_{k\geq j}\sum_{n=1}^l
\delta_{k \, i_n}.  \eee Then $B^{IJ}$ can be written in the following form
\bee\label{formB}
pB^{IJ}&=&\epsilon^{\alpha(I)\alpha(J)}\bigg(\sum_{i=1,j=1,i\neq j}^p \bigg(
   \nu_0 ( 1- (-1)^{i+j+\Delta(i)+\Delta(j)}) K_{ij} \nn &+& \nu_2 ( 1+
   (-1)^{i+j+\Delta(i)+\Delta(j)}) K_{ij}R_iR_j \bigg)       \nn &+&
 \nu_1\sum_{i=1}^p R_i\nn &+& \sum_{i=1}^p\sum_{j=p+1}^N (\nu_0 K_{ij} +\nu_2
 K_{ij}R_iR_j) \bigg), \eee and one can easily check using {\it Propositions
1 and 2} that  \\ {\it i)} $K_{ij}c(I)$ decomposes in the product of two even
cycles when $1\leq i <j\leq p$ and $i+j+\Delta(i)+\Delta(j)$ is odd;   \\
{\it ii)} $K_{ij}R_iR_j c(I)$ decomposes in the product of two even cycles
when $1\leq i <j\leq p$ and $i+j+\Delta(i)+\Delta(j)$ is even;  \\ {\it iii)}
$R_i c(I)$ is even cycle when $1\leq i \leq p$ because $c(I)$ is odd cycle;
\\ {\it iv)} if $c(K)\neq c(I)$ is some cycle in the decomposition
(\ref{dec}) of $g$ then $K_{ij}c(I)c(K)$ is the cycle with the same parity as
$c(K)$ has when $i \in C(I)$ and $j\in C(K)$;                          \\
{\it v)} if $c(K)\neq c(I)$ is some cycle in the decomposition (\ref{dec}) of
$g$ then $K_{ij}R_iR_jc(I)c(K)$ is the cycle with the same parity as $c(K)$
has when $i \in C(I)$ and $j\in C(K)$.

The case $c(I)\neq c(J)$ reduces to the subcases {\it iv)} and {\it v)}
considered above what ends the proof of {\it Lemma 1.}

\section{The supertraces on ${\fr G}$, Ground Level Conditions and the number
of supertraces on $H_{W(\R)}(\nu)$.} \label{Q_N}

Due to the $G$-invariance the definition of the supertrace on $\G$ is the
definition of the central function on $\G$ $i.e.$ a function on the conjugacy
classes of $G$ and so the number of the supertraces on $\G$ is equal to the
number of the conjugacy classes in $G$.

Since $\G \subset H_G(\nu)$ some additional restrictions on these functions
follow from (\ref{scom}) and the defining relations (\ref{KK})-(\ref{A}) of
$H_G(\nu )$.  Actually, consider some elements $b^I$ such that
$\lambda_I=-1$.  Then, one finds from (\ref{scom}) and (\ref{eigs}) that $str
\left ( b^I b^J g \right )$= $ - str \left ( b^J g b^I\right )$= $ str \left
( b^J b^I g \right )$ and therefore $ str \left ( [ b^I, b^{J}] g \right )
=0\,$ or equivalently \bee\label{GLC} \delta_{c(I)\,c(J)}
\delta_{-1\,\lambda_J}str(g)=-str(B^{IJ}g).  \eee Since these conditions
restrict supertraces of degree-0 polynomials of $a^\alpha_i$ we called them
in \cite{kv} as ground level conditions ({\it GLC}).  Due to {\it Lemma 2}
proved in Appendix the equations (\ref{GLC}) become identities for every
supertrace on ${\fr G}$ if $\delta_{c(I)\,c(J)} \delta_{-1\,\lambda_J}$=$0$.
Hence they express the supertrace of elements $g$ with $E(g)=e$ via the
supertraces of elements $B^{IJ}g$ with $E(B^{IJ}g)=e-1$:  \bee\label{GLC2}
str(g)=-str(B^{I}g), \eee where $B^{I}$ is $B^{IJ}$ with $J$ defined by
relations $c_J$ = $c_I$, $\lambda_I=\lambda_J=-1$, $\alpha(I)+
\alpha(J)=1$.  So the number of solutions $Q(G)$ of system (\ref{GLC}) does
not exceed ${\cal O}(G)$, the number  of conjugacy classes of elements
without odd cycles.

In the Appendix the following theorem is proved:

{\bf Theorem 2.} {\it  If $G=W(\R)$ where $\R=A_{N-1}$, $B_N$, $C_N$, $D_N$,
then  $\quad Q(G)={\cal O}(G)$}.

To compute the value ${\cal O}(G)$ consider some even cycle \bee\label{cyc}
c=K_{12} K_{23} \,...\,K_{(p-1)p} R_{i_1} R_{i_2}\,...\, R_{i_l}, \mbox {
($p+l$ is odd)} \eee and the sequence of similarity transformations
admissible in all cases \footnote{$R_i$-s are absent in the case
$\R=A_{N-1}$.} $\R=B_N, C_N, D_N$:  \bee\label{sim} c\rightarrow
R_{i_l}R_{i_l-1}cR_{i_l}R_{i_l-1}= K_{12} K_{23} \,...\,K_{(p-1)p} R_{i_1}
R_{i_2}\,...\, R_{i_l-2} \rightarrow ... \\ \rightarrow ... = K_{12} K_{23}
\,...\,K_{(p-1)p} (R_{1})^{l_1}(R_{2})^{l-l_1}.  \eee Here $l_1$ is the
number of odd $i_k$ in (\ref{cyc}).  If $p$ is even then either $l_1$ or
$l-l_1$ is odd and (\ref{cyc}) is similar to $K_{12} K_{23} \,...\,K_{(p-1)p}
R_{1}$.  If $p$ is odd then (\ref{cyc}) is similar either to $K_{12} K_{23}
\,...\,K_{(p-1)p}$ or to $K_{12} K_{23} \,...\,K_{(p-1)p} R_1 R_2$.  The
latter expression is similar to $K_{12} K_{23} \,...\,K_{(p-1)p} R_1 R_p$
that in its turn is similar to $K_{12} K_{23} \,...\,K_{(p-1)p}$ since $p$ is
odd.

This consideration shows that the conjugacy class of some even cycle $c(I)$
is determined by two dependent values: the number of indices $p(I)=|C(I)|$
that transformed by $c(I)$ and the parity
$\epsilon(I)=\left(p(I)+1\right)_{mod\,2}$  of the number of $R$-s in
$c(I)$.

In such a way the conjugacy class with $E=0$ completely characterized by the
set of nonnegative integers $n_1, \,n_3, \,...;\, m_2,\,m_4,\,...$, where
$n_i$ is the number of cycles $c(I)$ with $p(I)=i$ and $\epsilon(I)=0$ while
$m_i$ is the number of cycles $c(I)$ with $p(I)=i$ and $\epsilon(I)=1$.

The numbers $n_i$ and $m_i$ have to satisfy the following conditions:\\ for
the case  {\bf A)} $\quad m_i=0\quad \forall i\,,\qquad \sum_i \,in_i =N$;\\
for the case  {\bf B,C)} $\quad \sum_i \,i(n_i+m_i) =N\,$;                \\
for the case  {\bf D)}  $\quad \sum_i \,i(n_i+m_i) =N\,,\qquad \left (\sum_i
m_i\right )_{mod\,2}=0\, $.

So the number of supertraces is equal to a number of partitions of $N$ into a
sum of odd positive integers for the case $A_{N-1}$, a number of partitions
of $N$ into a sum of positive integers for the cases $B_N$ and $C_N$ and
$q(D_N)$ is a number of partitions of $N$ into a sum of positive integers
with even number of  even integers.

\vskip 5 mm
\vskip 5 mm
\noindent {\bf Acknowledgments} \vskip 3 mm
\noindent
Author is very grateful to M.~A.~Vasiliev for useful discussions.  This work
was supported in part by the Russian Basic Research Foundation, Grant
96-01-01144, Grant 96-02-17314, and INTAS Grant 93-0633.

\appendix

\vskip 8 mm \noindent {\large \bf Appendix}

\setcounter{equation}{0}
\renewcommand{\theequation}{Ap\arabic{equation}}
\section*{The proof of {\it Theorem 2}.}\label{app1}
First let us prove the following lemma.\\ {\it{\bf Lemma 2.\label{lemma2}}
The equations $str([b^I,\,b^J]g)$ with $b^I,b^J \in {\fr B}_g$,
$\lambda_I=-1$ are satisfied equivalently for any supertrace $str(\cdot)$ on
$\G$ if $c(I)\neq c(J)$ or $c(I)= c(J)$ and $\lambda_J\neq -1$.}

\noindent Indeed, due to $G$-invariance the following identities can be
obtained:  \bee &{}&str([b^I,\,b^J]g)=str(c(I)[b^I,\,b^J]g(c(I))^{-1})=\nn
&{}& str([c(I)b^I(c(I))^{-1},\,c(I)b^J(c(I))^{-1}]c(I)g(c(I))^{-1})=
str([-b^I,\,b^J]g)                 \nonumber \eee for the case $c(I)\neq
c(J)$, and $$ str([b^I,\,b^J]g)=str(c(I)[b^I,\,b^J]g(c(I))^{-1}) =-\lambda_J
str([b^I,\,b^J]g) $$ for the case $c(I)= c(J)$.

As a result one can consider only the case $c(I)=c(J)$, $\lambda_I=\lambda_J
=-1$, and $\alpha(I)=1-\alpha(J)$ to prove {\it Theorem 2}.

By induction on a number of odd cycles $e=E(g)$ we show that for $g$ with
$E(g)=e \geq 1$ there is only one independent equation on $str(g)$ provided
that all equations (\ref{GLC}) with $E(g)=e^\prime <e$ are resolved.  The
first step of the induction consists of the observation that there are no
equations for the case $E(g)=0$.

Let us consider the case where there are two equations (\ref{GLC}) on
$str(g)$ for some $g$. This is only possible if $g= c_1 c_2 g^\prime$ where
$c_1$ and $c_2$ are some odd cycles in the decomposition of $g$ such that
$c_1$ is not similar to $c_2$.  Note that $E(g^\prime)=E(g)-2=e-2$, $E(c_1
g^\prime)=E(c_2 g^\prime)=e-1$.

Without loss of generality let us set \bee\label{c1} c_1=K_{12} K_{23}
\,...\,K_{(p-1)p} R_{i_1} R_{i_2}\,...\, R_{i_k} \eee with $1\leq
i_1<i_2<\,...\,<i_k\leq p$ and odd $p-1+k$, and \bee\label{c2}
c_2=K_{(p+1)(p+2)} \,...\,K_{(p+q-1)(p+q)} R_{i_{k+1}}R_{i_{k+2}}\,...\,
R_{i_{k+l}} \eee with $p+1\leq i_{k+1}<i_{k+2}<\,...\, <i_{k+l} \leq p+q$ and
odd $q-1+l$, and introduce the corresponding generating elements
\bee\label{b1} b^\alpha_1= \frac 1 {\sqrt p}\sum_{s=1}^p (-1)^s
(-1)^{\Delta(s)} a^\alpha_s,\\ \label{b2} b^\alpha_2= \frac 1 {\sqrt
q}\sum_{s=p+1}^{p+q} (-1)^s (-1)^{\Delta(s)} a^\alpha_s,\nn \mbox{ where }
\Delta(s)=\sum_{u=s}^{p+q} \delta(u), \mbox{ and } \delta(u)=\sum_{v=1}^{k+l}
\delta_{ui_v}\,, \eee with eigenvalues $\lambda_1=\lambda_2=-1$.

Let us note that $c_+=K_{1\,(p+1)}c_1c_2$ is a cycle such that \bee\label{c+}
c_+ b^\alpha_+=-b^\alpha_+c_+\,,\mbox{ where } b^\alpha_+=\frac{\sqrt p
b^\alpha_1+\sqrt q b^\alpha_2}{\sqrt{p+q}}\,, \eee while
$c_-=K_{1\,{p+1}}R_1R_{p+1}c_1c_2$ is a cycle such that \bee\label{c-} c_-
b^\alpha_-=-b^\alpha_-c_-\,,\mbox{ where } b^\alpha_-=\frac{\sqrt p
b^\alpha_1-\sqrt q b^\alpha_2}{\sqrt{p+q}} \,.  \eee

Now consider the equation for $str(g) $ in the form \bee str(g)=-str(([
b^0_1,\,b^1_1]-1)g)=\nn -str\left( \bigg([ b^0_1,\,b^1_1]-1 -\frac 1 p
      \sum_{i=1}^p\sum_{j=p+1}^{p+q}(\nu_0 K_{ij}+\nu_2K_{ij}R_iR_j) \bigg) g
    \right) \label{induc}\\ -str\left( \frac 1 p
             \sum_{i=1}^p\sum_{j=p+1}^{p+q}(\nu_0 K_{ij}+ \nu_2K_{ij}R_iR_j)g
    \right) \,.  \nonumber \eee

Denote \bee h=\left(\left([ b^0_1,\,b^1_2]-1 -\frac 1 p
\sum_{i=1}^p\sum_{j=p+1}^{p+q}(\nu_0 K_{ij}+\nu_2K_{ij}R_iR_j)
\right)g\right),\qquad h\in{\fr G}, \nonumber \eee and note that $E(h)=e-1$,
and $h=\sum_t \gamma_t g_t$ with some $g_t\in G$ and constants $\gamma_t$,
such that every $g_t$ contains the cycle $c_2$ in its decomposition
(\ref{dec}).  So due to inductive hypothesis the following identity is true
$$ str(h)=-str(([ b^0_2,\,b^1_2]-1)h) $$ and we obtain \bee\label{1} str(g)=
str\left(\left([ b^0_2,\,b^1_2]-1\right)\left([ b^0_1,\,b^1_1]-1 -\frac 1 p
\sum_{i=1}^p\sum_{j=p+1}^{p+q}(\nu_0 K_{ij}+\nu_2K_{ij}R_iR_j)
\right)g\right) \nn -str(\frac 1 p \sum_{i=1}^p\sum_{j=p+1}^{p+q}(\nu_0
K_{ij}+\nu_2K_{ij}R_iR_j)g).  \eee The substitution $1\leftrightarrow 2$,
$p\leftrightarrow q$  gives \bee\label{2} str(g)= str\left(\left([
b^0_1,\,b^1_1]-1\right)\left([ b^0_2,\,b^1_2]-1 -\frac 1 q
\sum_{i=1}^p\sum_{j=p+1}^{p+q}(\nu_0 K_{ij}+\nu_2K_{ij}R_iR_j)
\right)g\right) \nn -str(\frac 1 q \sum_{i=1}^p\sum_{j=p+1}^{p+q}(\nu_0
K_{ij}+\nu_2K_{ij}R_iR_j)g).  \eee Let us show that due to inductive
hypothesis the difference between (\ref{1}) and (\ref{2}) is vanishing.  This
difference is equal to \bee \label{dif} X= str\left(\left(\frac{[
b^0_2,\,b^1_2]} p -\frac{[ b^0_1,\,b^1_1]} q \right)
\sum_{i=1}^p\sum_{j=p+1}^{p+q}(\nu_0 K_{ij}+\nu_2K_{ij}R_iR_j) g\right)\,,
\eee because $str([[b_1^0,\,b_1^1],\, [b_1^0,\,b_1^1]]g)=0$ for every
supertrace on ${\fr G}$.  Now we can use the following identities for $1<
i\leq p$ and $p+1< j \leq p+q$ \bee\label{id} (c_1)^{-1} K_{ij}
c_1=K_{(i-1)\,j}(R_{i-1}R_j)^{\delta(i-1)},\nn (c_2)^{-1} K_{ij}
c_2=K_{i\,j-1}(R_{i}R_{j-1})^{\delta(j-1)},\nn (c_1)^{-1} K_{ij}
R_iR_jc_1=K_{(i-1)\,j}(R_{i-1}R_j)^{\delta(i-1)+1},\nn (c_2)^{-1} K_{ij}
R_iR_jc_2=K_{i\,j-1}(R_{i}R_{j-1})^{\delta(j-1)+1} \nonumber \eee to deduce
that $X=X_0+X_2$ for every $G$-invariant supertrace on ${\fr G}$, where
\bee\label{diff} X_0= F_0(\nu_0,\nu_2) str\left(\left(\frac{[ b^0_2,\,b^1_2]}
p -\frac{[ b^0_1,\,b^1_1]} q \right) K_{1\,(p+1)}  g\right)\nn X_2=
F_2(\nu_0,\nu_2) str\left(\left(\frac{[ b^0_2,\,b^1_2]} p -\frac{[
b^0_1,\,b^1_1]} q \right) K_{1\,(p+1)} R_1R_{p+1} g\right) \eee and
$F_{0,\,1}(\nu_0,\,\nu_2)$ are some definite functions.  Substituting
$b^\alpha_2= \frac 1 {\sqrt q}(\sqrt{p+q}B^\alpha_+ - \sqrt p b^\alpha_1)$ in
$X_0$ and $b^\alpha_2= \frac 1 {\sqrt q}(\sqrt{p+q}B^\alpha_- + \sqrt p
b^\alpha_1)$ in $X_2$, and using inductive hypothesis and {\it Lemma 2} one
obtains $X_0=X_2=0$, what finishes the proof of {\it Theorem 2}.

\end{document}